\documentclass[doublecol]{epl2}
\usepackage{epsfig,epstopdf}
\usepackage{amsmath}
\usepackage{amssymb}
\usepackage{subfigure}
\usepackage{float}
\usepackage{cite}
\usepackage{hyperref}
\usepackage{cleveref}
\usepackage{graphicx}
\usepackage{graphics}

\title{Graphene with wedge disclination in the presence of intrinsic and Rashba spin orbit couplings}
\shorttitle{Bound state in graphene with wedge disclination} 

\author{Tarun Choudhari \and Nivedita Deo}
\shortauthor{T.Choudhari, N.Deo}

\institute{Departement of Physics and Astrophysics, University of Delhi.Delhi-110007,India.}

\pacs{72.10.Fk}{Scattering by point defects, dislocations, surfaces, and other imperfections}
\pacs{71.10.Pm}{Fermions in reduced dimensions}
\pacs{73.22.Pr}{Electronic structure of graphene}

\abstract{In this article, the  modified Kane-Mele Hamiltonian is derived for graphene with wedge disclination and spin orbit couplings (intrinsic and Rashba). The wedge disclination changes the flat lattice into the conical lattice and  hence modifies the spin orbit couplings. The Hamiltonian is exactly solved for the intrinsic spin orbit interaction and perturbatively for the Rashba spin orbit interaction. It is shown that there exists the Kramer's degenerate midgap localized spin separated fluxon states around the defect. These zero energy spin separated states occur at the external magnetic flux value $\Phi\pm\Delta\Phi$. The external magnetic flux $\Phi$ is introduced to make the wave-function periodic when the electron circulates around the defect. It is found that this separation occurs due to the effect of the conical curvature on the spin orbit coupling. Further, we find these results are robust to the addition of the Rashba spin orbit interaction which is important for the application to spintronics and nanoelectronics.}

\begin{document}

\maketitle
\section{Introduction\label{sec:1}}
"Topological Insulator (TI)" like HgTe, Graphene and Fluorinated Stanene\cite{B,D}, are materials which are insulators in bulk and have protected conducting gapless surface (3D) or edge (2D) states for electrons. This defines a new class of quantum materials called quantum spin hall insulators. These topological insulators are classified in terms of the symmetry, some of them preserve inversion and time reversal symmetry\cite{G,J}, while some of them have broken time reversal symmetry\cite{L,M}. The TI which preserves time reversal symmetry arises due to the presence of intrinsic spin orbit coupling. The robust properties of surface and edge states of topological insulators are important for applications to quantum computers and spintronics devices hence many studies have been made in the past few years to find new topological materials and structures. The lattice with topological defects is one of them. It has been naturally\cite{N}, as well as experimentally\cite{y,z}, and theoretically seen\cite{P,R,V,W,AB,BC}, that 3D topological insulators (TI) and topological superconductors (TS) show significant response to 'Topological Defects' and have gapless states around the defect for an electron circulating it. Recently the bound state of conical singularity in graphene with wedge disclination and broken time reversal symmetry have been studied for the spinless electron\cite{Y}, which shows the existence of localized zero energy bound states near the defect. The effect of Coulomb impurities in the presence of a uniform magnetic field on the energy spectrum of disclinated graphene has also been studied recently\cite{XZ}. The present paper deals with the bound state of the electron having spin in graphene with wedge disclination in the presence of curvature modified intrinsic and Rashba spin orbit coupling, in order to shed light on how the spin orbit coupling affects the bound state in disclinated graphene. The analysis has been done by using the Kane-Mele Hamiltonian\cite{J} for a spinfull electron on the honeycomb lattice. Although the Kane-Mele model (i.e.quantum spin hall insulator (QSHI) ) is difficult to realize in graphene because the intrinsic spin orbit coupling\cite{m} in pristine graphene is too weak, about 20-50 $\mu ev$, to have QSHI phase in it. This can be made large enough by having graphene absorbed with hydrogen ad-atoms\cite{v},  heavy elements\cite{w} or by proximity with $Sb_2Te_3$ TI\cite{x}. These methods increases the intrinsic SOC to the order of 20 mev to realize QSHI. In this paper we show that there exists the Kramer's degenerate midgap localized spin fluxon states\cite{Z,a} around the defect in graphene with wedge disclination, and these zero energy up and down spin states separate out purely due to the effect of the curvature of the conical honeycomb lattice on the intrinsic spin orbit coupling in the presence of the external magnetic flux. So in this work we first introduce the wedge disclination, derive the curvature modified intrinsic spin orbit coupling Hamiltonian and hence the Kane-Mele Hamiltonian for the conical honeycomb lattice. We then find the bound state wave function and energy of the low energy electron in the presence of intrinsic and Rashba SOC.
\section{Wedge disclination and gauge transformation due to wedge disclination}
\begin{figure}
\centering
\includegraphics[width=45mm,height=38mm]{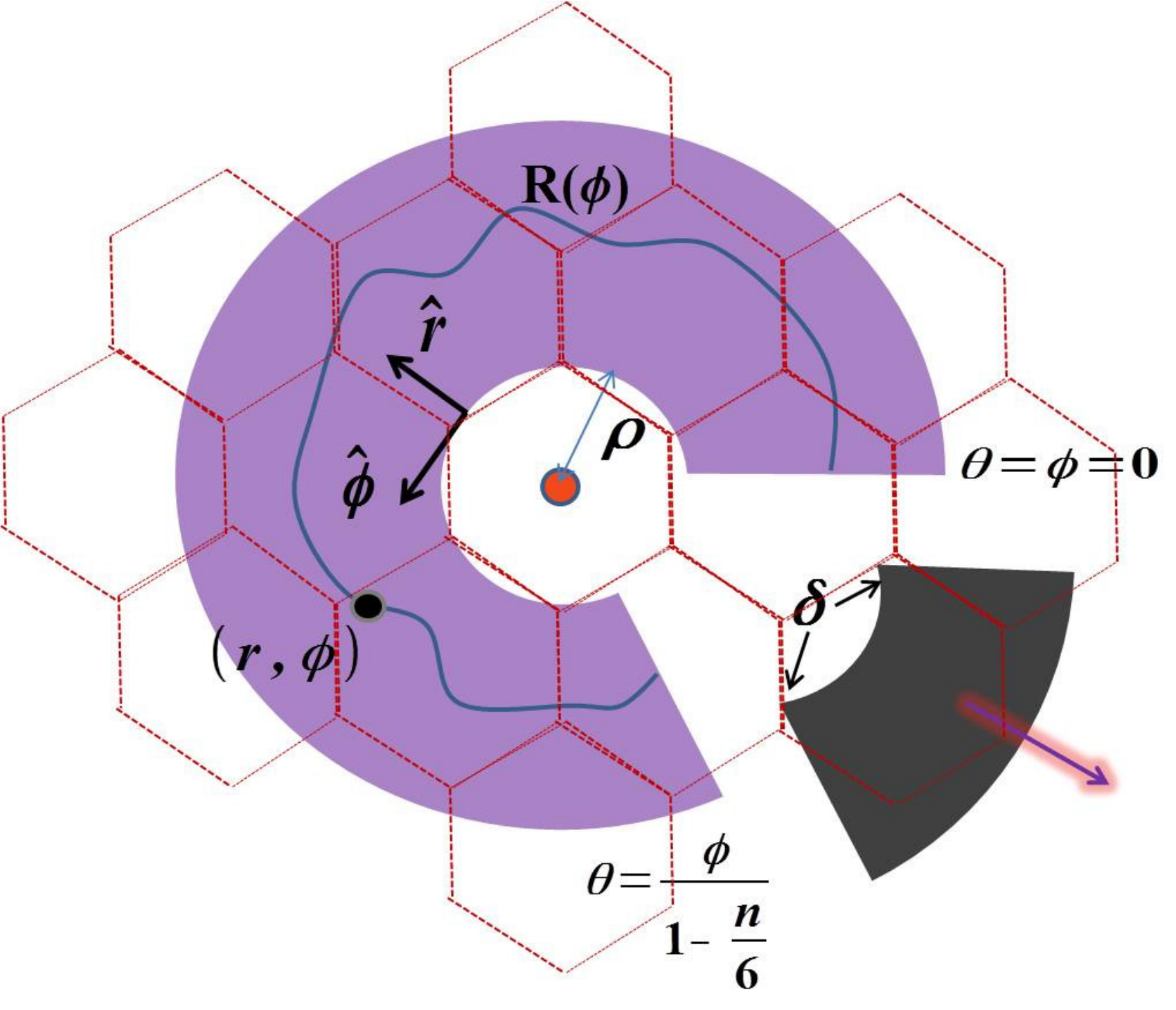}
\caption{(color online)Unfolded plane of lattice (Continuum model) with wedge of angle $\delta$ removed from it. $\rho$ is the radius of the defect hole. $R(\phi)$ is the curve along which the spinor is rotated. At the centre the red dot represent the flux tube with magnetic flux $\Phi$. $(r,\phi)$ are the coordinates of a point on the lattice.}
\label{fig:continum}
\end{figure}

\begin{figure}
\centering
\includegraphics[width=75mm,height=35mm]{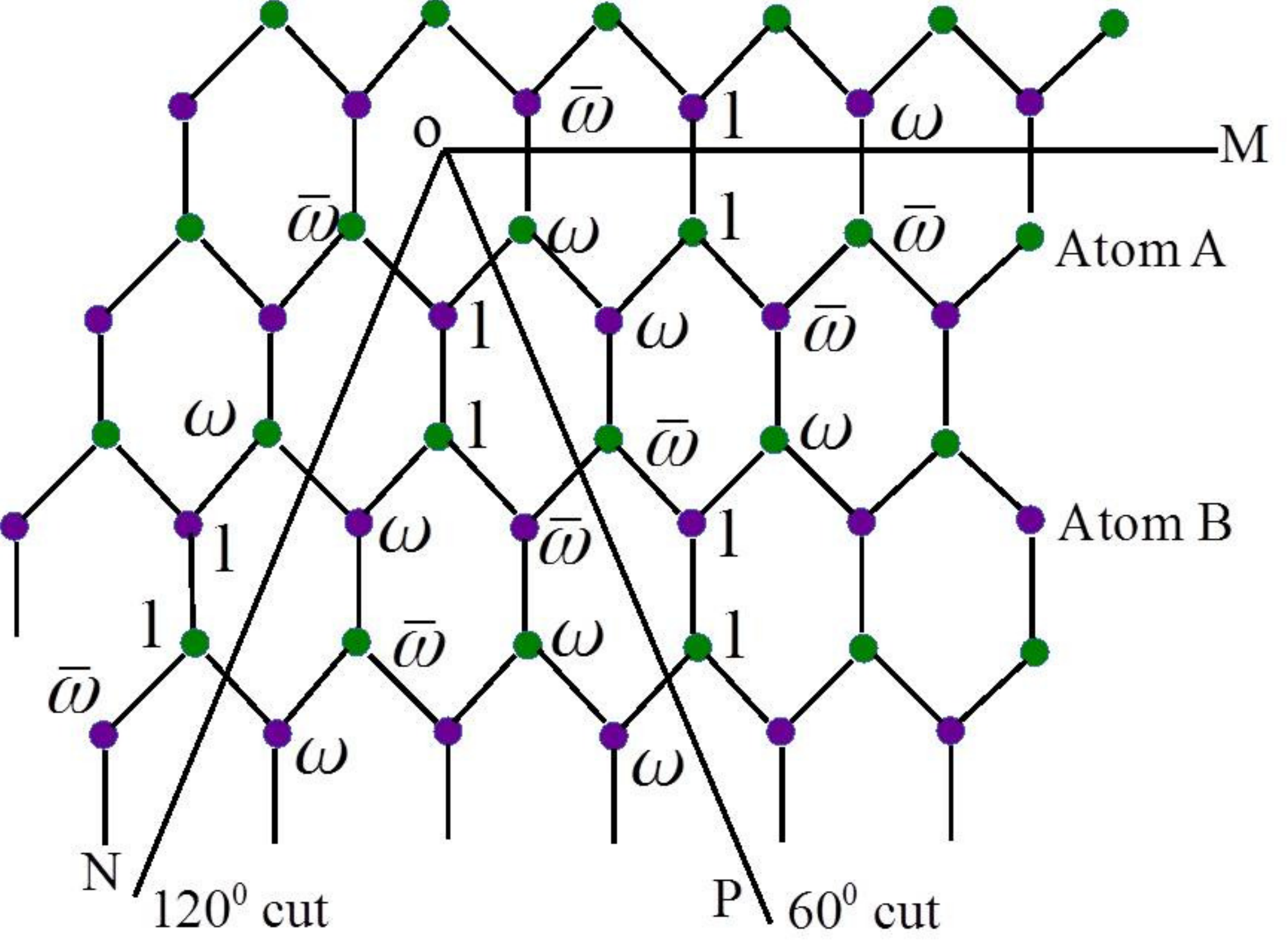}
\caption{(color online)The figure shows the distribution of the phases ($\omega=\mathrm{e}^{\mathrm{i}2\pi/3}$ and $\bar{\omega}=\mathrm{e}^{-\mathrm{i}2\pi/3}$) of base function $u_{A\uparrow}$ and $u_{B'\uparrow}$ on the honeycomb lattice.For base functions $u_{A\downarrow}$ and $u_{B'\downarrow}$ the distribution of phases is same.For the $60^o$ and $120^o$ cut the phases of base functions are matched along the line(OM and OP) and (OM and ON) respectively. The distribution of the phases ($\omega$ and $\bar{\omega}$) for base function ($u_{A'\downarrow}$,$u_{B\downarrow}$) and  ($u_{A'\uparrow}$,$u_{B\uparrow}$) is just the complex conjugate of the above figure.}
\label{fig:phase1}
\end{figure}

The low energy spinfull electron dynamics in flat honeycomb lattice is governed by the massive 2D Dirac Hamiltonian, called the Kane-Mele model for honeycomb lattice \cite{J}, and is given by
\begin{equation}
H(\mathbf{k})=\nu_{\mathrm{f}}\hbar(\tau_{\mathrm{z}}\sigma_{\mathrm{x}}k_{\mathrm{x}}+\sigma_{\mathrm{y}}k_{\mathrm{y}})+\Delta_{\mathrm{so}} \tau_{\mathrm{z}}\sigma_{\mathrm{z}}s_{\mathrm{z}},
\label{eq:1}
\end{equation} 
where the $\mathbf{\tau}$, $\mathbf{\sigma}$ and \textbf{s} are the Pauli matrices for valley, sublattice and spin space. $\tau_{\mathrm{z}} =\pm 1$ for the two valley K(K') in the Brioullin zone, $\sigma_{\mathrm{z}} =\pm1$ for the sublattice A(B), $s_{\mathrm{z}}=\pm 1$ for up and down spin of the electron. $\nu_f$ is the Fermi velocity  and $\Delta_{\mathrm{so}}=3\sqrt{3}t_2$ defines the intrinsic spin orbit coupling parameter in the honeycomb lattice, $t_2$ defines the second nearest neighbour hoping amplitude. The total low energy electron wave function for this Hamiltonian is $\Psi(r)=\left[(u_{A\uparrow},u_{A\downarrow}, u_{B\uparrow},u_{B\downarrow}),(u_{A'\uparrow},u_{A'\downarrow},u_{B'\uparrow},u_{B'\downarrow})\right]\times\psi(r)$. Where $u_{(A,B,A',B')(\uparrow\downarrow)}$(r) are the base functions for the sublattice A(B) for momentum K, and for momentum K', for up and down spin of the electron. $\psi(r)$ is a 8x1 envelope wave function slowly varying on the lattice. The Kane-Mele Hamiltonian above [eq. (\ref{eq:1})] has an intrinsic spin orbit coupling Hamiltonian as the mass term, which produces a gap in the energy spectrum and is given by, $H_{\mathrm{so}}=\Delta_{\mathrm{so}}\tau_{\mathrm{z}}\sigma_{\mathrm{z}}s_{\mathrm{z}}$, this Hamiltonian preserves time reversal symmetry. In graphene, the wedge disclination is define by a procedure called the Volterra's cut and glue procedure in which there is a removal or addition of the wedge of angle $n\pi/3$ from the flat lattice followed by the gluing of the edges of a cut to form a continuous lattice by preserving the three fold connectivity of the atoms. The resultant lattice then has a conical shape with apex as a hole in the form of a polygon. The index 'n' defines the type of disclination, for $n>0$ the wedge is removed (positive curvature) and for $n<0$ there is an addition of the wedge (negative curvature) in the flat honeycomb lattice. So there are four types of disclination in a honeycomb lattice which are given by the pentagon defect (n=1), the square defect (n=2), the heptagon defect (n=-1) and the octagon defect (n=-2). Graphene in nano cone form have defects with more than one pentagon and heptagon at the top of the cone. It has been found experimentally\cite{y,z} that these nano cones can have a single pentagon cell at the top of the cone as well which leads to the existence of the wedge disclination in graphene. To study the bound states of the electron around these defects we have used the continuum model, which introduces the curvature induced gauge field in the low energy electron Dirac Hamiltonian [eq. (\ref{eq:1})] of the honeycomb lattice \cite{R,AB,BC}. In the continuum model, the honeycomb lattice with disclination is considered as an unfolded plane (shown in fig. \ref{fig:continum}) with a cut of angle $\delta$ and the hole  of radius '$\rho$' (located in place of the polygon at the top of cone) such that any point on this continuum lattice is defined by the polar coordinates (r,$\phi$).
So in the Volterra's cut and glue procedure for the disclinated honeycomb lattice, the wavefunction $\Psi$ for the electron with spin must be single valued at the seam if this spinor is rotated around the defect core along the curve $R(\phi)$ (fig. \ref{fig:continum}). This leads to the general gauge transformation for the envelope function $\psi$ so as to compensate for the mismatch of phases of the base functions  $u_{A\uparrow}\:, u_{A\downarrow}\:,u_{B\uparrow}\:, u_{B\downarrow}\:,u_{A'\uparrow} \:,u_{A'\downarrow} \:,
u_{B'\uparrow}\:,u_{B'\downarrow}$. The Bloch phases are given by  $\omega=\mathrm{e}^{\mathrm{i}2\pi/3}$ and $\bar{\omega}=\mathrm{e}^{-\mathrm{i}2\pi/3}$. In fig. \ref{fig:phase1} the distribution of  phases for these base functions is shown.
By matching the phases for any  general disclination (n),\cite{Y} the general gauge transformation, \cite{AB,BC} for electrons with spin is given by
\begin{equation}
\psi(\phi=2\pi-\frac{n\pi}{3})=e^{i\frac{n\pi}{6}(\tau_{\mathrm{z}}\sigma_{\mathrm{z}}s_o-3\tau_{\mathrm{y}}\sigma_{\mathrm{y}}s_o)}\psi(\phi=0),
\label{eq:3}
\end{equation}
where the $s_o$ is the 2x2 identity matrix for the spin space. The cut and glue procedure transforms the flat honeycomb lattice into a conical honeycomb lattice. So if we rescale the angle $\phi$ of unfolded plane to the new angle\, $\theta=\phi/(1-n/6)$\, where $\theta$ is varying from 0 to $2\pi$ so as to have conical topology in the unfolded plane then the general gauge transformation for any given disclination defined by index, $n=\pm 1,\pm 2 $  is
\begin{equation}
\psi(\theta=2\pi)=e^{i\frac{n\pi}{6}(\tau_{\mathrm{z}}\sigma_{\mathrm{z}}s_o-3\tau_{\mathrm{y}}\sigma_{\mathrm{y}}s_0)}\psi(\theta=0).
\label{eq:4}
\end{equation}
This transformation can be broken into two singular transformations\cite{Y}, for the envelope wave function ($\psi$) of the electron with spin so that $\psi(\theta=2\pi)=U_{\mathrm{s}}(\phi)V_{n\mathrm{s}}(\theta)\psi(\theta=0)$,
where
\begin{equation}
U_{\mathrm{s}}(\phi)=e^{i\frac{\phi}{2}\tau_{\mathrm{z}}\sigma_{\mathrm{z}} s_o},\quad V_{n\mathrm{s}}(\theta)=e^{i\frac{n\theta}{4}\tau_{\mathrm{y}}\sigma_{\mathrm{y}}s_o}.
\label{eq:5}
\end{equation}
\section{Kane-Mele Hamiltonian for the conical honeycomb lattice}
Now consider an external magnetic flux $\Phi$ applied at the centre of the defect hole using magnetic flux tube (red dot in fig. \ref{fig:continum}) having a vector potential $\textbf{A}$ given by 
 \begin{equation}
 \textbf{A}=\frac{\Phi}{r\Omega_n\Phi_0}\hat{\phi}.
 \end{equation}
with $\Omega_n=(1-n/6)$ \,and  $\Phi_0 = h/e$ the magnetic flux quanta. To define the dynamics for the electron with spin in the honeycomb lattice with disclination we have used the Kane-Mele model [eq. (\ref{eq:1})] at low energy in polar coordinates $(r,\theta)$ with a external magnetic flux $\Phi$. However, in this Kane-Mele model the intrinsic spin orbit coupling Hamiltonian ($H_{\mathrm{so}}=\Delta_{\mathrm{so}}\tau_{\mathrm{z}}\sigma_{\mathrm{z}}s_{\mathrm{z}}$) is changed into the curvature modified intrinsic spin orbit coupling ($H'_{\mathrm{so}}$) for the conical honeycomb lattice because the lattice curvature strongly affects the spin orbit coupling\cite{m}. So if we consider the honeycomb lattice as a cone having semi-cone angle $\alpha$, then the coordinates of a point on its surface is defined by $(r,\theta,z)$. We derived $H'_{\mathrm{so}}$, using the transformation $\Lambda$ obtained by transforming the spin basis of the perpendicular component of the spin ($s_{\mathrm{z}}$) on the flat honeycomb lattice to the spin basis of the perpendicular component of the spin ($s_n$) on the conical lattice\cite{m}. The transformation is given by
\begin{equation}
\Lambda=\frac{1}{\sqrt{2}}\left(\begin{array}{cc} e^{-i\frac{\theta}{2}}\sqrt{1+\sin{\alpha}} & e^{i\frac{\theta}{2}}\sqrt{1-\sin{\alpha}}\\e^{-i\frac{\theta}{2}}\sqrt{1-\sin{\alpha}} & -e^{i\frac{\theta}{2}}\sqrt{1+\sin{\alpha}} \\\end{array} \right),
\end{equation}
The Hamiltonian  $H'_{\mathrm{so}}=\Lambda^{\dagger}H_{\mathrm{so}}\Lambda$ is then found to be
\begin{multline}
H'_{\mathrm{so}}=\Delta_{\mathrm{so}}\tau_{\mathrm{z}}\sigma_{\mathrm{z}} s_{\mathrm{z}}f(\alpha)+\Delta_{\mathrm{so}}\tau_{\mathrm{z}}\sigma_{\mathrm{z}}s_{\mathrm{x}}\cos\theta g(\alpha)\\-\Delta_{\mathrm{so}}\tau_{\mathrm{z}}\sigma_{\mathrm{z}}s_{\mathrm{y}} \sin\theta g(\alpha),
\label{eq:8}  
\end{multline} 
where $g(\alpha)=\cos{\alpha}$ and $f(\alpha)=\sin{\alpha}$, are the curvature functions and they depend upon the type of wedge disclination (n) through the semicone angle $\alpha$, $\sin \alpha=1-|n|/6$\cite{l}. It is interesting to note that this curvature modified intrinsic SOC Hamiltonian has two extra spin flipping intrinsic Rashba like terms $\tau_{\mathrm{z}}\sigma_{\mathrm{z}}s_{\mathrm{x}}$ and $\tau_{\mathrm{z}}\sigma_{\mathrm{z}} s_{\mathrm{y}}$ which occur due to the mixing of $p_z$ with $p_x$ and $p_y$ orbitals of the carbon atoms present at the curved surface of the conical graphene. This is in contrast to the the intrinsic SOC Hamiltonian $H_{\mathrm{so}}$ for the flat honeycomb lattice and the cylindrical nano tube SOC but is similar to the spherical honeycomb lattice SOC Hamiltonian\cite{m}. The Kane-Mele Hamiltonian in polar coordinates is then given by
\begin{equation}
H_{\mathrm{FS}}(\textbf{k})=\nu_f\hbar(\tau_{\mathrm{z}}\sigma_{\mathrm{x}}k_{r}+\sigma_{\mathrm{y}} k_{\theta})+\frac{\Phi}{r\Omega_n\Phi_0}\sigma_{\mathrm{y}}+ H'_{\mathrm{so}},
\label{eq:10}
\end{equation}
where $k_{r}=-\mathrm{i}\partial/\partial r$ and $k_{\theta}=-\mathrm{i}/r\Omega_n(\partial/\partial\theta)$.
Now by using the Hamiltonian $ H'_{\mathrm{so}}$ [eq. (\ref{eq:8})] in the Hamiltonian $H_{\mathrm{FS}}$ [eq. (\ref{eq:10})] and by applying the gauge transformation $U_{\mathrm{s}}(\phi)$ and $V_{n\mathrm{s}}(\theta)$ on the Hamiltonian $H_{\mathrm{FS}}$, we find the Hamiltonian for the spinfull electron in the honeycomb lattice with wedge disclination to be given by $H_{\mathrm{DS}}(r,\theta)=U_{\mathrm{s}}^{\dagger}V_{n\mathrm{s}}^{\dagger}H_{\mathrm{FS}}V_{n\mathrm{s}}U_{\mathrm{s}}$,
\begin{multline}
H_{\mathrm{DS}}=\left[k_{r}-\frac{i}{2r}\right]\tau_{\mathrm{z}}\sigma_{\mathrm{x}}+\left[k_{\theta}+ \frac{\Phi}{r\Omega_n\Phi_0}+\frac{n}{4r\Omega_n}\tau_{\mathrm{z}}\right]\sigma_{\mathrm{y}}\\\\+\Delta_{\mathrm{so}}\left[\tau_{\mathrm{z}}\sigma_{\mathrm{z}}s_{\mathrm{z}} f(\alpha)
+\tau_{\mathrm{z}}\sigma_{\mathrm{z}} s_{\mathrm{x}}g(\alpha)\right],
\label{eq:11}  
\end{multline}
where $\hbar=\nu_f=1$. We have used $H'_{\mathrm{so}}$ with the condition $\theta=2\pi$ because $\theta=\phi/(1-n/6)$\, and for $\phi=2\pi-\delta$, $\theta$ is always $2\pi$. The curvature of conical graphene manifests itself in the generalized Kane-Mele Hamiltonian through the gauge potential like term: $i/2r$ and $n/4r\Omega_n$ occurring due to the gauge transformations $U_{\mathrm{s}}(\phi)$ and $V_{n\mathrm{s}}(\theta)$ and through the functions $f(\alpha)$ and $g(\alpha)$ occurring in the spin orbit coupling mass term ($\Delta_{\mathrm{so}}$). In spite of the presence of Rashba like term in this Hamiltonian an exact solution can be found where as for spherical and spiral honeycomb lattice this is not the case\cite{m},\cite{FD}.

\section{The bound state wavefunction and energy for the electron around the defect}
 The Hamiltonian $(H_{DS})$, is separable in the coordinates 'r' and '$\theta$'. By using the ansatz $\psi(r,\theta)=e^{ij\theta}X(r)$ where 'j' is a half integer number and acts as the azimuthal angular momentum quantum number for the electron circulating around the defect core, we find the modified Kane-Mele Hamiltonian in radial coordinates to be given by
\begin{multline}
H'_{\mathrm{DS}}(r)=\left[\left(k_{r}-\frac{i}{2r}\right)\tau_{\mathrm{z}}\sigma_{\mathrm{x}}+ \frac{\nu_{\tau}}{r}\sigma_{\mathrm{y}}\right]\\+\Delta_{\mathrm{so}}\left[\tau_{\mathrm{z}}\sigma_{\mathrm{z}}s_{\mathrm{z}} f(\alpha)
+\tau_{\mathrm{z}}\sigma_{\mathrm{z}} s_{\mathrm{x}}g(\alpha)\right]
\label{eq:12}
\end{multline}  
where $\tau=\pm$ for the two emergent valleys and $\nu_{\tau}$ is
\begin{equation}
\nu_\tau=\frac{j+\frac{\Phi}{\Phi_0}+\frac{n\tau}{4}}{(1-\frac{n}{6})}.
\label{eq:13}
\end{equation}
If we solve the eigenvalue equation $H'_{\mathrm{DS}}X(r)=\epsilon X(r)$, we find the bound state spinor $\psi(r,\theta)$ for the electron with spin in terms of the modified Bessel functions of the second kind which decay exponentially for $r\rightarrow\infty$,
\begin{equation}
 \psi(r,\theta)=e^{ij\theta}\left[\begin{array}{c}K_{\nu_+ -\frac{1}{2}}(r\kappa)\\ \, \\ \frac{(i\kappa+\Delta_{\mathrm{so}}g(\alpha))}{(\epsilon+\Delta_{\mathrm{so}}f(\alpha))}K_{\nu_+ -\frac{1}{2}}(r\kappa)\\ \, \\ \frac{(i\kappa-\Delta_{\mathrm{so}}g(\alpha))}{(\epsilon+\Delta_{\mathrm{so}}f(\alpha))}K_{\nu_+ +\frac{1}{2}}(r\kappa)\\\, \\ K_{\nu_+ +\frac{1}{2}}(r\kappa) \\\,\\\frac{(-i\kappa-\Delta_{\mathrm{so}}g(\alpha))}{(\epsilon+\Delta_{\mathrm{so}}f(\alpha))}K_{\nu_- +\frac{1}{2}}(r\kappa)\\\, \\ K_{\nu_- +\frac{1}{2}}(r\kappa) \\\,\\ K_{\nu_- -\frac{1}{2}}(r\kappa) \\\,\\\frac{(-i\kappa+\Delta_{\mathrm{so}}g(\alpha))}{(\epsilon+\Delta_{\mathrm{so}}f(\alpha))}K_{\nu_- -\frac{1}{2}}(r\kappa)\\ \end{array}\right].
 \label{eq:14}
 \end{equation}
where $\kappa=\sqrt{(\Delta_{\mathrm{so}})^2-\epsilon^2}$. Now the square integrability of $\psi$ does not uniquely define the quantized state as $\rho\rightarrow0$.
So to find this, we have assumed a confining potential $V(r<\rho)=-m_0(\tau_o\sigma_{\mathrm{z}}s_o)$ in the region $r<\rho$ in place of the spin orbit coupling mass term in the Kane-Mele model [eq. (\ref{eq:1})]. Then by making $m_0\rightarrow\infty$ we have found the infinite mass boundary condition at r=$\rho$ for the spinfull electron given by $\psi(\rho,\theta)=M\psi(\rho,\theta)$, where \textbf{M} is a general 8x8 matrix. We have found the matrix \textbf{M} of infinite mass boundary condition for the spinfull electron by using the procedure given in \cite{o} for Dirac current operator, $\hat{J}=\tau_{\mathrm{z}}\sigma_{\mathrm{x}} \hat{r}+ \sigma_{\mathrm{y}} \hat{\phi}$ and the time reversal operator, $T=-i(\tau_{\mathrm{y}}\otimes I\otimes s_{\mathrm{y}})C,$ for the [eq. \ref{eq:1}]. The matrix 'M' is given by 
$M=\tau_{\mathrm{z}}\otimes\hat{\phi}.\mathbf{\sigma}\otimes s_{\mathrm{z}}.$ So the boundary condition becomes
\begin{equation}
\psi(\rho,\theta=2\pi)=\tau_{\mathrm{z}}\otimes\sigma_{\mathrm{y}}\otimes s_{\mathrm{z}}\psi(\rho,\theta=2\pi).
\label{eq:15}
\end{equation}
The boundary condition [eq. (\ref{eq:15})] leads to the bound state energy of the up and down spin electron at the K and K' valleys and is given by

\begin{multline}
\frac{\epsilon_{\gamma_z}}{\Delta_{\mathrm{so}}}=\frac{-1+\frac{K_{\nu_{\tau} +\gamma\frac{1}{2}}(\rho\kappa)}{K_{\nu_{\tau}-\gamma\frac{1}{2}}(\rho\kappa)}\sqrt{\frac{K^2_{\nu_{\tau} +\gamma\frac{1}{2}}(\rho\kappa)}{K^2_{\nu_{\tau} -\gamma\frac{1}{2}}(\rho\kappa)}R_{\alpha}+(R_{\alpha}-f^2 )}}{1+\frac{K^2_{\nu_{\tau} +\gamma\frac{1}{2}}(\rho\kappa)}{K^2_{\nu_{\tau} -\gamma\frac{1}{2}}(\rho\kappa)}}     
\label{eq:16}
\end{multline}  
where $\gamma=\pm1$ for upspin ($\gamma_z=\uparrow$) and down ($\gamma_z=\downarrow$) spin state respectively. $R_{\alpha}=[(g(\alpha))^2+1]$. These bound state energies depend upon the semi-cone angle $\alpha$, hole radius $\rho$ and the the parameter $\nu_{\tau}$. Further, if we put $f(\alpha)=1$ and $R(\alpha)=1$ (i.e., $g(\alpha)=0$) in the eq. (\ref{eq:16}), which means the mass term $\Delta_{\mathrm{so}}$ now becomes independent of the curvature, we find the same result found by Ruegg and Lin\cite{Y}, for the bound state energy of the spin-less electron,$
\sqrt{\Delta_{\mathrm{so}}-\epsilon}/\sqrt{\Delta_{\mathrm{so}}+\epsilon}=K_{\nu_\tau -\frac{1}{2}}(\rho\kappa)/K_{\nu_\tau +\frac{1}{2}}(\rho\kappa),$ where $\Delta_{\mathrm{so}}=m$ is the mass term of the Haldane Hamiltonian for the honeycomb lattice.
\section{Addition of Rashba Hamiltonian} We have also included the Rashba SOC to probe whether the zero energy localized states are robust against the addition of the disorder terms like the Rasbha spin orbit coupling which may arise due to the interaction with a substrate or an external electric field. The Rashba Hamiltonian\cite{m}, in the flat honeycomb lattice is given by $ H_R=\lambda_{\mathrm{R}}(\tau_{\mathrm{z}}\sigma_{\mathrm{x}} s_{\mathrm{y}}-\sigma_{\mathrm{y}}s_{\mathrm{x}})$. For the conical lattice the curvature modified Rashba Hamiltonian is found to be, $H'_R=\Lambda^{\dagger}H_R\Lambda,$ given by
\begin{multline}
H'_R=-\lambda_{\mathrm{R}}\left(\tau_{\mathrm{z}}\sigma_{\mathrm{x}} s_{\mathrm{y}} \cos{\theta} +\tau_{\mathrm{z}}\sigma_{\mathrm{x}}s_{\mathrm{x}} \sin{\theta}-\sigma_{\mathrm{y}}s_{\mathrm{z}}g(\alpha)\right)
\\+\lambda_{\mathrm{R}}\left(\sigma_{\mathrm{y}} s_{\mathrm{x}} f(\alpha) \cos{\theta}-\sigma_{\mathrm{y}}s_{\mathrm{y}} f(\alpha) \sin{\theta} \right).
\label{eq:17}
\end{multline}
Note that this curvature modified Rashba  SOC (which flips spins on hoping) contains a spin preserving term $\sigma_{\mathrm{y}} s_{\mathrm{z}}$ which occurs in this Hamiltonian purely due the curvature of the conical honeycomb lattice. Now we have included this modified Rashba Hamiltonian, $H'_{\mathrm{R}}$, in our Hamiltonian, $H_{\mathrm{DS}}$[eq. (\ref{eq:11})], for the honeycomb lattice with wedge disclination and the spin full electron after applying the gauge transformation $U_{\mathrm{s}}(\phi)$ and $V_{n\mathrm{s}}(\theta)$. The final Hamiltonian after addition of the modified Rashba Hamiltonian is given by, $H_{\mathrm{DSR}}= H_{\mathrm{DS}} + H'_R.$ We observe that this Hamiltonian is not solvable analytically. So we treat the Rashba Hamiltonian $H'_R$ as a perturbation in the Hamiltonian $H_{\mathrm{DS}}$ because $\lambda_R/\Delta_{\mathrm{so}}<1$ and used the degenerate perturbation theory to find the energy spectrum of the bound states for the Hamiltonian $H_{\mathrm{DSR}}$. The change in energy, $\Delta\epsilon$ due Rashba Hamiltonian is the eigenvalue of the W-matrix\cite{book}  and the total bound state energy is given by $
\epsilon=\large\epsilon_{\gamma_z}+\Delta\large\epsilon_\Gamma,$,where the $\epsilon_{\gamma_z}$ is the energy defined by eq. (\ref{eq:16}) and $\Gamma=\pm1$ defines the two energy states of mixed spin states around the defect. $\Delta\epsilon_{\Gamma}$ depends upon the curvature of the lattice through the function $f(\alpha)$ and $g(\alpha)$. Results after addition of Rashba SOC are discussed in the later part of the next section.
\section{Results and Discussion}
\begin{figure}
\vspace{-0.2cm}
\subfigure{\includegraphics[width=43mm,height=40mm]{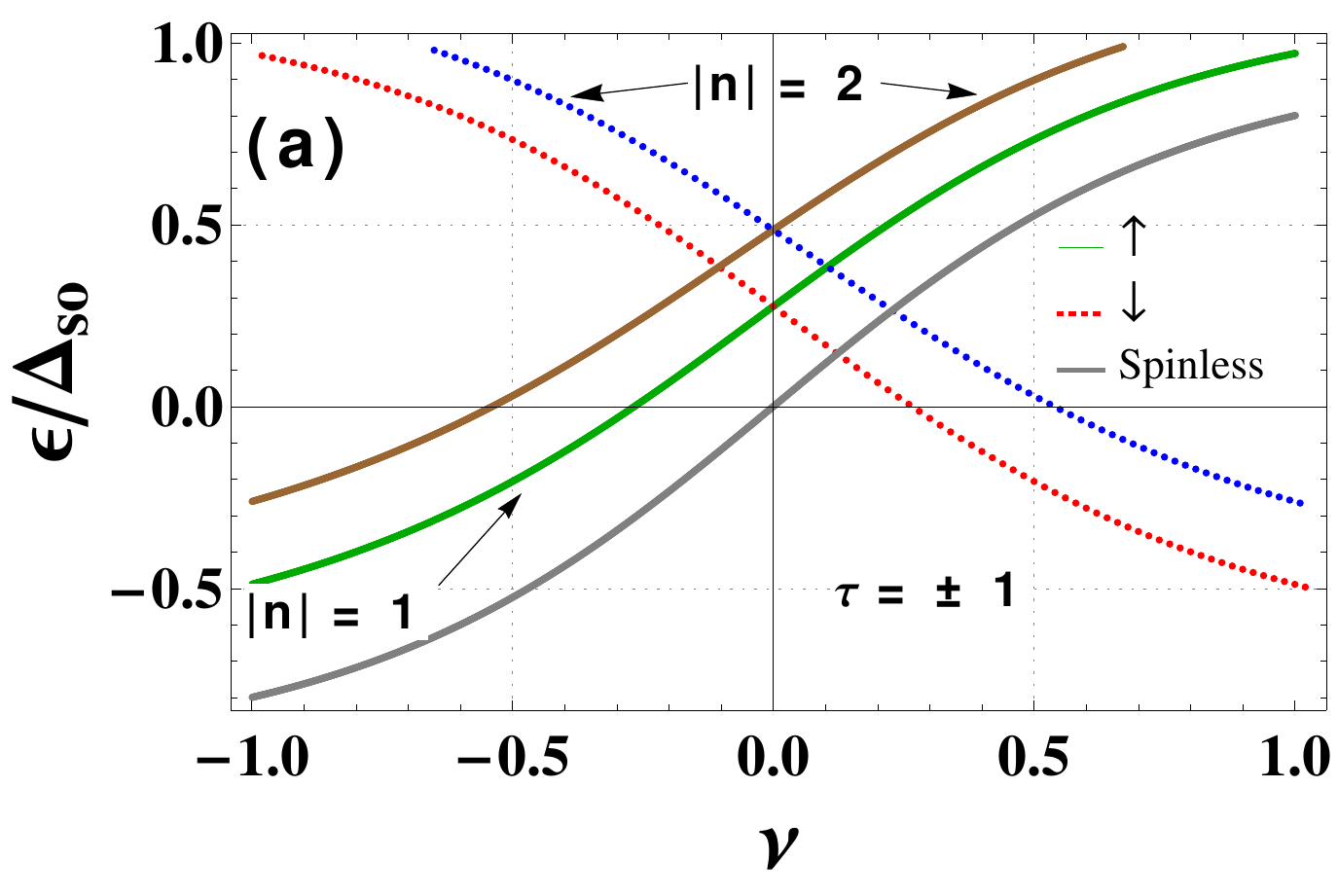}\label{fig:vnDeltaso}}
\subfigure{\includegraphics[width=43mm,height=40mm]{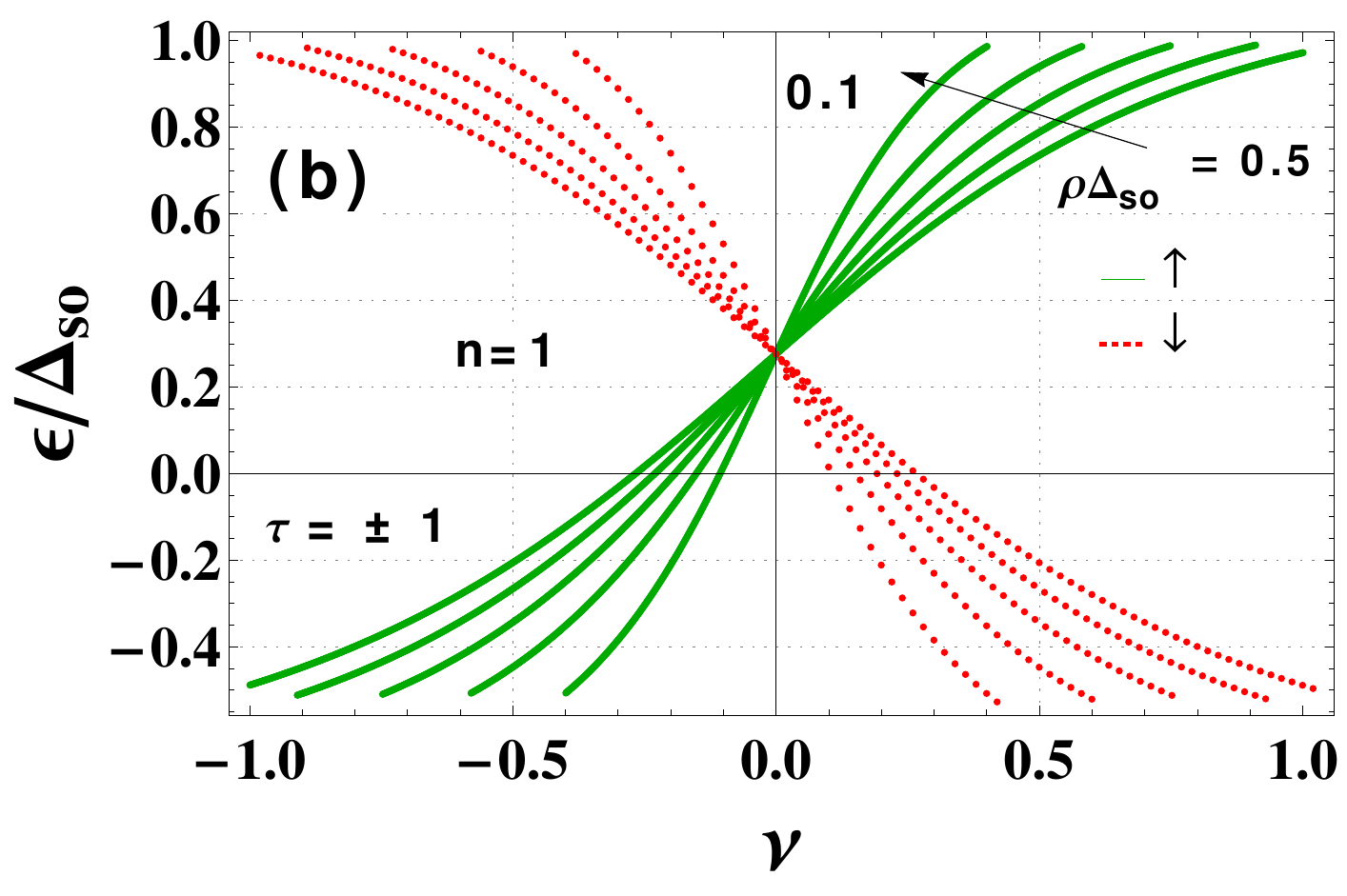}\label{fig:vdeltaso}}
 \subfigure{\includegraphics[width=42mm,height=38mm]{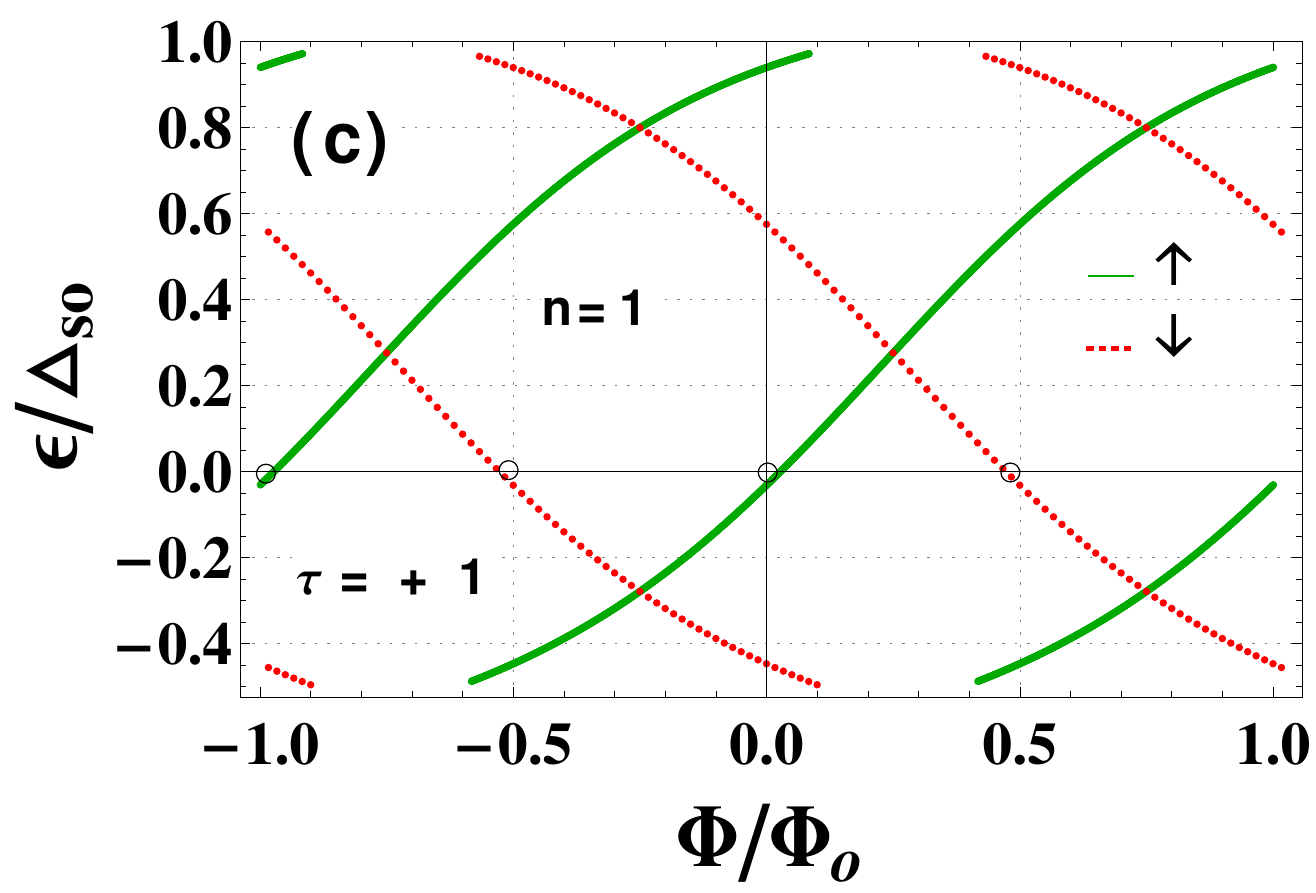}
\label{fig:EFluxn=1a}} 
\subfigure{\includegraphics[width=42mm,height=38mm]{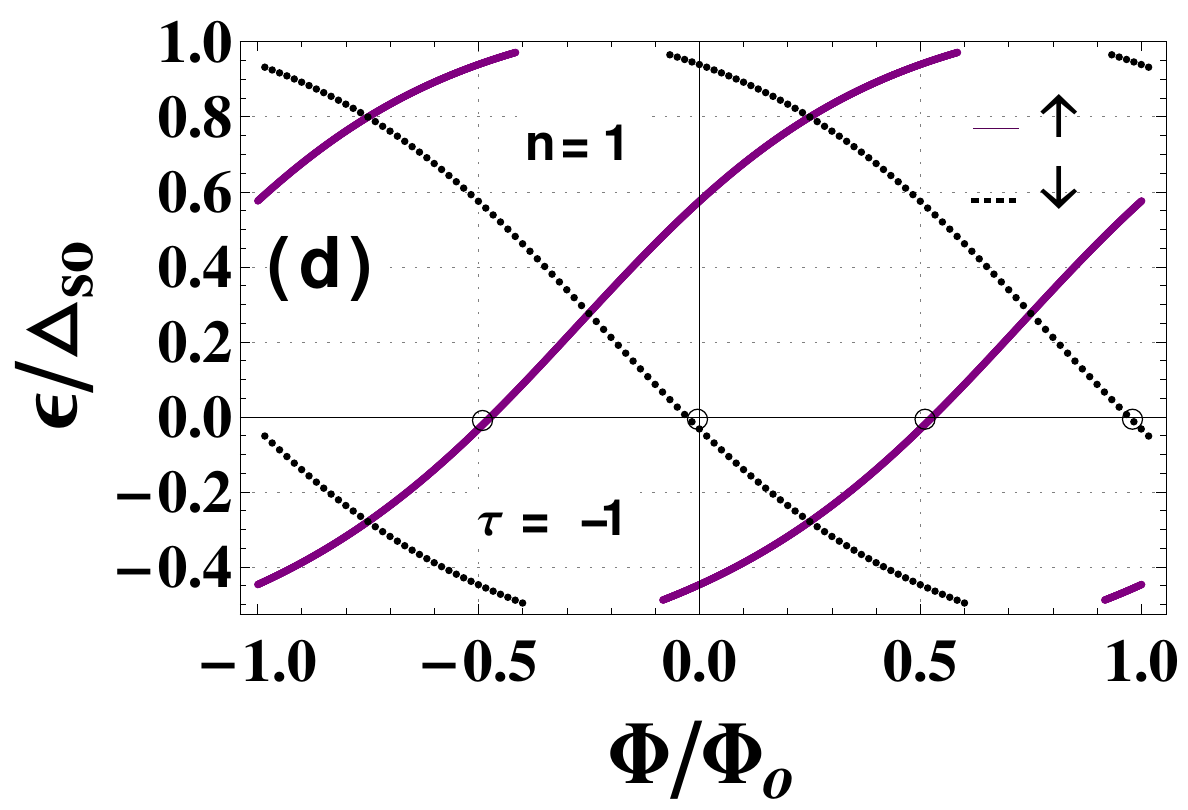}\label{fig:EFluxn=1b}}
\subfigure{\includegraphics[width=42mm,height=38mm]{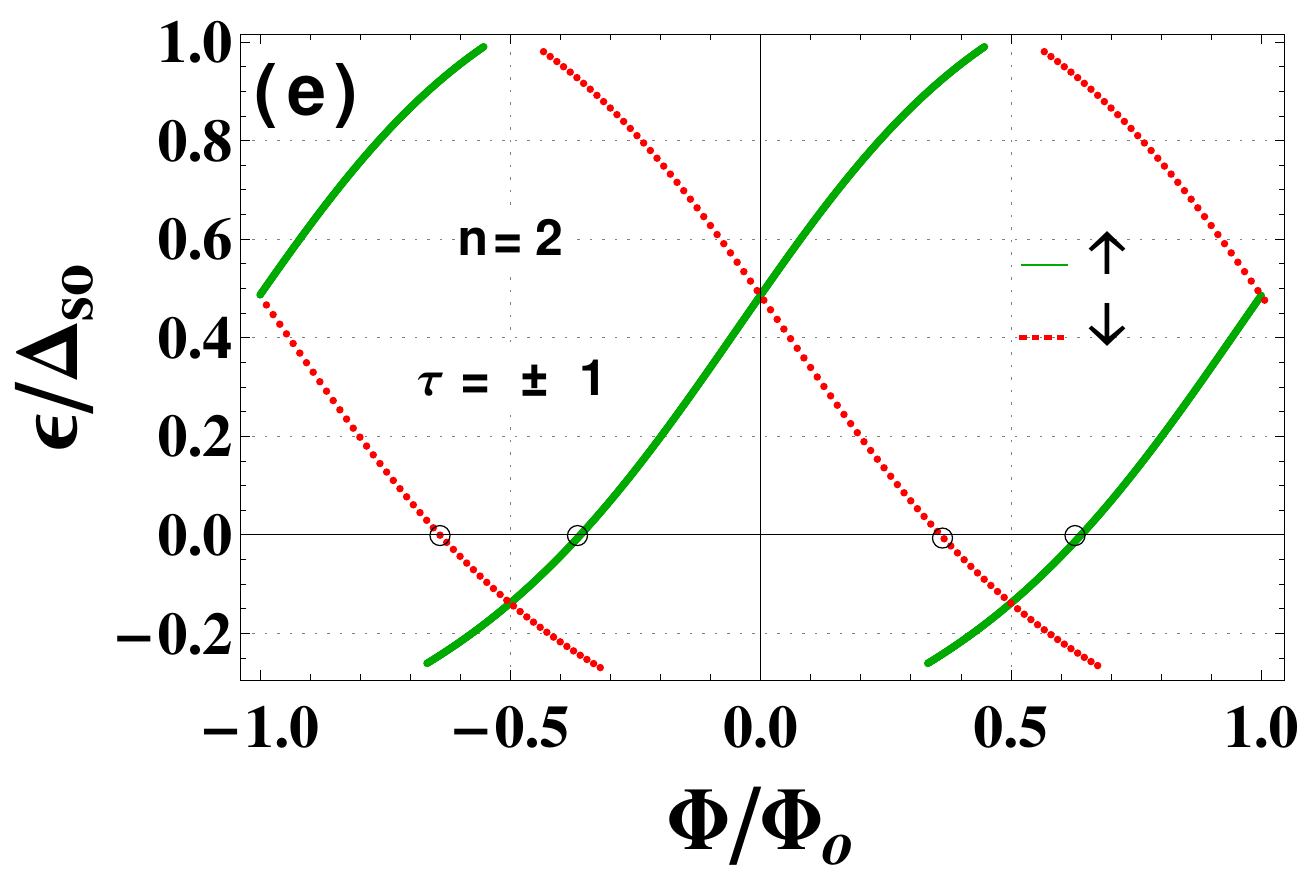}\label{fig:EFluxn=2ab}}
\subfigure{\includegraphics[width=43mm,height=38mm]{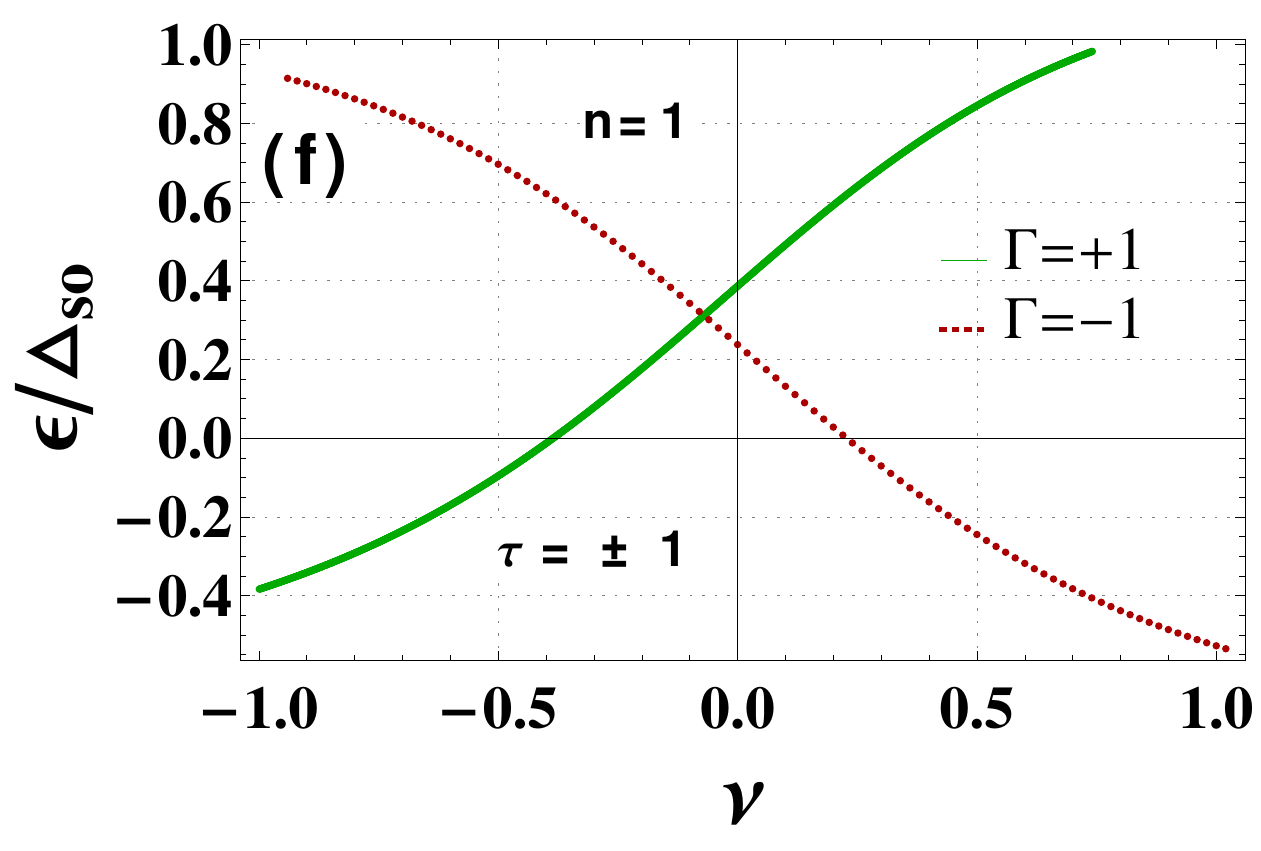}\label{fig:Rashba}}\vspace{-0.1cm}
\caption{(color online)(a),The general $\epsilon$ vs $\nu$ curve for upspin(Green,Brown,continuous) ,downspin (Red,Blue,dotted) for all four defects and spinless electron(Grey curve passing through origin) for pentagon defect(n=1) in two K valley for $\Delta_{\mathrm{so}}\neq0$ and $\lambda_R =0$. (b) $\epsilon$ vs $\nu$ curve for variable $\rho\Delta_{\mathrm{so}}$ ranging from 0.1 to 0.5. (c)-(e),$\epsilon$ vs $\Phi/\Phi_o$ plot for pentgon(n=1) and square (n=2) defect in two K valleys. (f),$\epsilon$ vs $\nu$ plot for pentgon defect(n=1) with $\Delta_{\mathrm{so}}\neq0$ and $\lambda_R\neq0$.All plots are for $\rho\Delta_{\mathrm{so}}=0.5$.}
\end{figure}

For $\Delta_{\mathrm{so}}\neq0$ and $\lambda=0$, the energy spectrum for the bound state ($\epsilon<\Delta_{\mathrm{so}}$) of the electron for up and down spin states for the pentagon (n=1), square (n=2), heptagon (n=-1) and octagon (n=-2) defect in the two valleys, K and K' is shown in fig. \ref{fig:vnDeltaso}, which we have plotted using eq. (\ref{eq:16}) and incorporating $\sin \alpha=1-|n|/6$, with respect to general $\nu$ (first and second valley). In the fig. \ref{fig:vnDeltaso} the continuous green curve shows the up spin energy and the dotted red curve shows the down spin state energy for pentagon defect (n=1). We find that the up and down spin state energy monotonically varies with $\nu$ with a crossing at $\nu=0$. The energy spectrum of the spin-less electron is also shown in fig. \ref{fig:vnDeltaso} as the grey curve crossing the origin. First, we conclude from the energy spectrum (fig. \ref{fig:vnDeltaso}), that the zero energy mode, which occurs at $\nu=0$ for spin-less electron are also present for up spin and down spin states but are now separated and are at non zero value of $\nu=\pm \nu_o$ respectively. Now the magnitude of $\nu_o$ depends upon the value of external magnetic flux ($\Phi$) and the type of disclination (n) and this separation purely occurs due to the effect of curvature on the intrinsic spin orbit coupling mass term $\Delta_{\mathrm{so}}$ through the curvature functions $g(\alpha)$ and $f(\alpha)$.
Second, from this plot we conclude that $\nu_{\tau\downarrow}=-\nu_{\tau\uparrow}$. Then in the expression for $\nu_{\tau}$ given by eq. (\ref{eq:13}) as $\nu_\tau=(j+\Phi/\Phi_0+n\tau/4)/\Omega_n$, the occurrence of the term $\Phi^f=n\tau/4$ in $\nu_{\tau}$ is due to the wedge disclination and is called the fictitious magnetic flux experienced by the electron when it circulates around the defect core. Then for a given spin $\gamma_z$=($\uparrow,\downarrow$), $\nu_{\tau \gamma_z}$ , $j_{\gamma_z}$ and external flux $\Phi$ the fictitious magnetic flux will be given by $\Phi^f_{\tau s_o}=\nu_{\tau s_o}\Omega_n-j_{s_o}-\Phi/\Phi_o$. Then for $\Phi=0$ and $j_{\uparrow}=-j_{\downarrow}$, we get $\Phi^f_{\tau\uparrow}=-\Phi^f_{\tau\downarrow}$,
which implies that the fictitious spin flux $(n\Phi_o/4)$ due to wedge disclinations has opposite signs for the up spin and down spin states. 
Third, the zero energy up spin and down spin states in the two valleys are localized around the defect as can be seen from the bound state spinor eq. (\ref{eq:14}) which decays exponentially with 'r' and the defect acts like a source of spin flux $(\tau \gamma_z n\Phi_o/4)$, we recognize these as the zero energy spin fluxon states, \cite{Z,a}. Fourth, from fig. \ref{fig:vnDeltaso} we observe that (i) for the $|n|=2$ the bound state energy is greater than the energy for $|n|=1$ (ii). the bound state energy of the positive curvature ($n > 0$) and the negative curvature ($n < 0$) are degenerate. The fig. \ref{fig:vnDeltaso} also shows the Kramer's degenerate pairs for up spin and down spin.
The fig. \ref{fig:vdeltaso} shows the energy spectrum of the bound state of the spinfull electron for different values of the product $\rho\Delta_{\mathrm{so}}$ of the hole radius $\rho$ and spin orbit coupling constant $\Delta_{\mathrm{so}}$ for the pentagon defect (n=1). This shows that at a given value of $\nu$ the bound state energy for both up and down spin decrease with increase in the value of $\rho\Delta_{\mathrm{so}}$ ranging from 0.1 to 0.5. However, for higher values of the product $\rho\Delta_{\mathrm{so}}$ the allowed range for $\nu$ to have bound state energy $\epsilon<\Delta_{\mathrm{so}}$ is increased. This shows that the  localized zero energy state around the defect still persists for a minimum size of the defect core ($\rho$) threaded by magnetic flux ($\Phi$). Further, the spectrum shows that the crossing of the up-spin and the down spin state remains fixed with a change in the magnitude of $\rho\Delta_{\mathrm{so}}$.
The fig. \ref{fig:EFluxn=1a}-\ref{fig:EFluxn=2ab} shows the bound state energy of the up spin and down spin states, at both K valleys, with respect to external magnetic flux ($\Phi$) for pentagon (n=1) and square defect(n=2). Now in graphene for the electron moving around the defect core, the spinor picks up a Berry phase of $\pi$ after rotation by angle $2\pi$ which leads to the antisymmetric boundary condition. So to get the periodic boundary condition and zero energy states we have to apply a external magnetic flux of $\Phi=h/2e=\Phi_o/2$ in the defect core. But as the defect also acts like the source of fictitious magnetic flux which is $\tau\,\Phi_o/4$ for pentagon defect (n=1) and $\tau\,\Phi_o/2$ for square defect (n=2), so to get zero energy bound state and periodic boundary condition we have to apply an external magnetic flux of $\Phi=\tau\,\Phi_o/4$ and $\Phi=0$ for the pentagon and square defect respectively as shown in $\epsilon$ vs $\Phi/\Phi_o$ plot of \cite{Y} for the spin less electron. But for the spinfull electron we find that the up and down spin zero energy states do not occur exactly at these value of external magnetic flux but they are separated and occur at the external magnetic flux, $\Phi \pm \Delta\Phi$. For the pentagon defect in the first valley K, the zero energy states for spin less electron occur at external magnetic flux value, $\Phi=-3\pi/2,\,\pi/2$ \cite{Y}, but for spinfull electron fig. \ref{fig:EFluxn=1a}, the pair of up and down spin zero energy states occur at external magnetic flux value $\Phi\pm\Delta\Phi=\pi/2\pm\pi/2$ and $\Phi\pm\Delta\Phi=-3\pi/2\pm\pi/2$. For the square defect (n=2), fig. \ref{fig:EFluxn=2ab} for which the value of the external magnetic flux should be $0$ or $\pm2\pi$ to get spin less zero energy states, we find the zero energy states occur at a external magnetic flux value $\Phi\pm\Delta\Phi=0 \pm3\pi/4$ and $\Phi\pm\Delta\Phi=-2\pi\pm 3\pi/4 \,,2\pi\pm 3\pi/4$. Similarly for the heptagon defect (n=-1) and for the octagon defect (n=-2) (will be reported elsewhere) the $\Delta\Phi$ is $7/5(\pi/2)$ and $3\pi/2$ respectively. This extra external magnetic flux, $\Delta\Phi$ occurs due to the effect of curvature on the intrinsic SOC through the curvature functions $g(\alpha)$ and $f(\alpha)$ because if we make the curvature function $g(\alpha)=0$ and $f(\alpha)=1$ (to remove the effect of spin), we find that the zero energy states occur at external magnetic flux value $\Phi$. So for all these defects the up spin zero energy state is present at the external magnetic flux value $\Phi-\Delta\Phi$ and the down spin zero energy state is present at the external magnetic flux value $\Phi+\Delta\Phi$ which concludes that they are separated. From fig. \ref{fig:EFluxn=1a}-\ref{fig:EFluxn=2ab}, we also observe the number of  up-spin and down-spin states for the pentagon (n=1), square defect(n=2), heptagon (n=-1) and octagon defect(n=-2) change from 2 to 1, 2 to 1, 3 to 2 and 3 to 2 respectively at different values of magnetic flux. Which means the external magnetic flux applied at the defect core can change the local density of states around the defect. Now for $\Delta_{\mathrm{so}},\lambda_{\mathrm{R}}\neq0$
, we have numerically plotted the energy spectrum of the Hamiltonian ($H_{\mathrm{DSR}}$) for the bound state electron given by $\epsilon=\epsilon_{\mathrm{DS}}+ \Delta\epsilon_\Gamma$ as shown in fig. \ref{fig:Rashba} for the first and second valley with  $\lambda_{\mathrm{R}}/\Delta_{\mathrm{so}}=0.3$ for the pentagon defect. Compared to the energy spectrum shown in the fig. \ref{fig:vnDeltaso}, for which the states of up and down spin cross each other at $\nu=0$ in the two K valleys, the states now cross each other at $\nu\neq0$. This change in $\nu$ for crossing depends upon the magnitude of the ratio of $\lambda_{\mathrm{R}}/\Delta_{\mathrm{so}}$. Further, the energy spectrum shows that the zero energy states are still present in the spectrum for both the valleys but now move to smaller values of $\nu=\pm\nu_r$ for the two K valleys as compared to fig. \ref{fig:vnDeltaso} in which they are at $\nu=\pm\nu_o$ such that $\nu_r<\nu_o$. In the plot for the bound state energy vs the external magnetic flux($\Phi$) for all four defects, we observe that the value of $\Delta\Phi$ after adding the Rashba spin orbit coupling does not change and it is still given by $\Delta\Phi$ equal to $\pi/2$, $3\pi/4$, $7/5(\pi/2)$ and $3\pi/2$ for pentagon (n=1), square(n=2), heptagon (n=-1) and octagon defect(n=-2) respectively.

\section{Conclusion}
In summary, we have found the bound state of the spin-full electron in  graphene with wedge disclination using the modified Kane-Mele model in the presence of curvature modified, intrinsic spin orbit coupling ($\Delta_{\mathrm{so}}$) and extrinsic Rashba spin orbit coupling ($\lambda_R$), for the conical graphene lattice. This study leads to the conclusion that for the case $\Delta_{\mathrm{so}}\neq 0$ and $\lambda_R=0$, for the electron circulating the defect, there exist Kramer's degenerate pair of zero energy spin bound states localized around the defect, acting as the source of spin fictitious flux and these zero energy spin states separates only due to curvature modified SOC of the wedge disclinated graphene in the presence of external magnetic flux. For the case  $\Delta_{\mathrm{so}}\neq 0$ and $\lambda_R\neq 0$ these zero energy spin states still persist and all other results are robust. This concludes that in 2D, disclinated graphene based topological insulators can act as a source of protected localized zero energy separated spin states around the defect which can be controlled by external magnetic flux. Hence these results can have potential applications in spintronics and nanoelectronics.
\acknowledgments
TC would like to thank the University Grant Commission (\textbf{U.G.C}) for Senior research fellowship (\textbf{S.R.F}). ND would like to thank the University of Delhi R \& D Research Grant.

\end{document}